\documentclass[twocolumn]{aastex631}
\shortauthors{Palma et al.}
\begin{document}

%\title{Template \aastex Article with Examples: 
%v6.3.1\footnote{Released on March, 1st, 2021}}

\title{Gamma-Ray Burst observations by the high-energy charged particle detector on board the CSES-01 satellite between 2019 and 2021}

\correspondingauthor{Francesco Palma}
\email{francesco.palma@roma2.infn.it}

\author[0000-0001-7076-8830]{F. PALMA}
\affiliation{INFN-Sezione di Roma ``Tor Vergata'', Via della Ricerca Scientifica 1, I-00133 Roma, Italy}

\author[0000-0002-3033-4824]{M. MARTUCCI}
\affiliation{INFN-Sezione di Roma ``Tor Vergata'', Via della Ricerca Scientifica 1, I-00133 Roma, Italy}

\author[0000-0002-2008-8404]{C. NEUB\"USER}
\affiliation{INFN-TIFPA, Via Sommarive 14, I-38123 Povo (Trento), Italy}

\author[0000-0001-8835-2796]{A. SOTGIU}
\affiliation{INFN-Sezione di Roma ``Tor Vergata'', Via della Ricerca Scientifica 1, I-00133 Roma, Italy}

\author[0000-0003-2317-9560]{F.~M. FOLLEGA}
\affiliation{University of Trento, Via Sommarive 14, I-38123 Povo (Trento), Italy}
\affiliation{INFN-TIFPA, Via Sommarive 14, I-38123 Povo (Trento), Italy}

\author[0000-0003-0601-0261]{P. UBERTINI}
\affiliation{INAF-IAPS, Via Fosso del Cavaliere 100, I-00133, Roma, Italy}

\author[0000-0002-2017-4396]{A. BAZZANO}
\affiliation{INAF-IAPS, Via Fosso del Cavaliere 100, I-00133, Roma, Italy}

\author[0000-0003-2126-5908]{J. RODI}
\affiliation{INAF-IAPS, Via Fosso del Cavaliere 100, I-00133, Roma, Italy}

\author[0000-0003-4501-3289]{R. AMMENDOLA}
\affiliation{INFN-Sezione di Roma ``Tor Vergata'', Via della Ricerca Scientifica 1, I-00133 Roma, Italy}

\author[0000-0002-9027-2039]{D. BADONI}
\affiliation{INFN-Sezione di Roma ``Tor Vergata'', Via della Ricerca Scientifica 1, I-00133 Roma, Italy}

\author[0000-0002-3066-8621]{S. BARTOCCI}
\affiliation{INFN-Sezione di Roma ``Tor Vergata'', Via della Ricerca Scientifica 1, I-00133 Roma, Italy}

\author[0000-0002-5808-7239]{R. BATTISTON}
\affiliation{University of Trento, Via Sommarive 14, I-38123 Povo (Trento), Italy}
\affiliation{INFN-TIFPA, Via Sommarive 14, I-38123 Povo (Trento), Italy}

\author[0000-0003-4673-8038]{S. BEOL\`E}
\affiliation{University of Torino, Via P. Giuria 1, I-10125 Torino, Italy}
\affiliation{INFN-Sezione di Torino, Via P. Giuria 1, I-10125 Torino, Italy}

\author[0000-0002-5260-416X]{I. BERTELLO}
\affiliation{INAF-IAPS, Via Fosso del Cavaliere 100, I-00133, Roma, Italy}

\author[0000-0003-1977-6354]{W.~J. BURGER}
\affiliation{INFN-TIFPA, Via Sommarive 14, I-38123 Povo (Trento), Italy}

\author[0000-0003-1504-9707]{D. CAMPANA}
\affiliation{INFN-Sezione di Napoli, Via Cintia, I-80126, Napoli, Italy}

\author[0000-0002-8107-9624]{A. CICONE}
\affiliation{University of L'Aquila, Via Vetoio, I-67100, L'Aquila, Italy}

\author{P. CIPOLLONE}
\affiliation{INFN-Sezione di Roma ``Tor Vergata'', Via della Ricerca Scientifica 1, I-00133 Roma, Italy}

\author[0000-0001-7470-4463]{S. COLI}
\affiliation{INFN-Sezione di Torino, Via P. Giuria 1, I-10125 Torino, Italy}

\author[0000-0003-2966-2000]{L. CONTI}
\affiliation{Uninettuno University, Corso V. Emanuele II, 39, I-00186, Roma, Italy}
\affiliation{INFN-Sezione di Roma ``Tor Vergata'', Via della Ricerca Scientifica 1, I-00133 Roma, Italy}

\author[0000-0002-2535-5700]{A. CONTIN}
\affiliation{University of Bologna, Viale Berti Pichat 6/2, Bologna, Italy}
\affiliation{INFN-Sezione di Bologna, Viale Berti Pichat 6/2, Bologna, Italy}

\author[0000-0002-0127-1342]{M. CRISTOFORETTI}
\affiliation{Fondazione Bruno Kessler, Via Sommarive 18, I-38123 Povo (Trento), Italy}
\affiliation{INFN-TIFPA, Via Sommarive 14, I-38123 Povo (Trento), Italy}

\author[0000-0002-9214-2051]{G. D'ANGELO}
\affiliation{INAF-IAPS, Via Fosso del Cavaliere 100, I-00133, Roma, Italy}

\author{F. DE ANGELIS}
\affiliation{INAF-IAPS, Via Fosso del Cavaliere 100, I-00133, Roma, Italy}

\author[0000-0002-9725-1281]{C. DE DONATO}
\affiliation{INFN-Sezione di Roma ``Tor Vergata'', Via della Ricerca Scientifica 1, I-00133 Roma, Italy}

\author[0000-0002-7280-2446]{C. DE SANTIS}
\affiliation{INFN-Sezione di Roma ``Tor Vergata'', Via della Ricerca Scientifica 1, I-00133 Roma, Italy}

\author[0000-0001-8279-020X]{P. DIEGO}
\affiliation{INAF-IAPS, Via Fosso del Cavaliere 100, I-00133, Roma, Italy}

\author{A. DI LUCA}
\affiliation{Fondazione Bruno Kessler, Via Sommarive 18, I-38123 Povo (Trento), Italy}
\affiliation{INFN-TIFPA, Via Sommarive 14, I-38123 Povo (Trento), Italy}

\author{E. FIORENZA}
\affiliation{INAF-IAPS, Via Fosso del Cavaliere 100, I-00133, Roma, Italy}

\author[0000-0001-7252-7416]{G. GEBBIA}
\affiliation{University of Trento, Via Sommarive 14, I-38123 Povo (Trento), Italy}
\affiliation{INFN-TIFPA, Via Sommarive 14, I-38123 Povo (Trento), Italy}

\author[0000-0001-5038-2762]{R. IUPPA}
\affiliation{University of Trento, Via Sommarive 14, I-38123 Povo (Trento), Italy}
\affiliation{INFN-TIFPA, Via Sommarive 14, I-38123 Povo (Trento), Italy}

\author{A. LEGA}
\affiliation{University of Trento, Via Sommarive 14, I-38123 Povo (Trento), Italy}
\affiliation{INFN-TIFPA, Via Sommarive 14, I-38123 Povo (Trento), Italy}

\author{M. LOLLI}
\affiliation{INFN-Sezione di Bologna, Viale Berti Pichat 6/2, Bologna, Italy}

\author{B. MARTINO}
\affiliation{CNR, Via Fosso del Cavaliere 100, I-00133, Roma, Italy}

\author[0000-0002-8911-1561]{G. MASCIANTONIO}
\affiliation{INFN-Sezione di Roma ``Tor Vergata'', Via della Ricerca Scientifica 1, I-00133 Roma, Italy}

\author{M. MERG\`E}
\affiliation{Italian Space Agency, Via del Politecnico, I-00133 Roma, Italy}
\affiliation{INFN-Sezione di Roma ``Tor Vergata'', Via della Ricerca Scientifica 1, I-00133 Roma, Italy}

\author{M. MESE}
\affiliation{University of Napoli ``Federico II'', Via Cintia, I-80126, Napoli, Italy}
\affiliation{INFN-Sezione di Napoli, Via Cintia, I-80126, Napoli, Italy}

\author{A. MORBIDINI}
\affiliation{INAF-IAPS, Via Fosso del Cavaliere 100, I-00133, Roma, Italy}

\author[0000-0002-4355-7947]{F. NOZZOLI}
\affiliation{INFN-TIFPA, Via Sommarive 14, I-38123 Povo (Trento), Italy}

\author{F. NUCCILLI}
\affiliation{INAF-IAPS, Via Fosso del Cavaliere 100, I-00133, Roma, Italy}

\author[0000-0002-6612-6170]{A. OLIVA}
\affiliation{INFN-Sezione di Bologna, Viale Berti Pichat 6/2, Bologna, Italy}

\author[0000-0002-9871-8103]{G. OSTERIA}
\affiliation{INFN-Sezione di Napoli, Via Cintia, I-80126, Napoli, Italy}

\author[0000-0003-3707-0013]{F. PALMONARI}
\affiliation{University of Bologna, Viale Berti Pichat 6/2, Bologna, Italy}
\affiliation{INFN-Sezione di Bologna, Viale Berti Pichat 6/2, Bologna, Italy}

\author[0000-0003-1063-6961]{B. PANICO}
\affiliation{University of Napoli ``Federico II'', Via Cintia, I-80126, Napoli, Italy}
\affiliation{INFN-Sezione di Napoli, Via Cintia, I-80126, Napoli, Italy}

\author[0000-0002-7969-7415]{E. PAPINI}
\affiliation{INAF-IAPS, Via Fosso del Cavaliere 100, I-00133, Roma, Italy}

\author[0000-0002-9073-3288]{A. PARMENTIER}
\affiliation{INAF-IAPS, Via Fosso del Cavaliere 100, I-00133, Roma, Italy}

\author[0000-0003-2868-2819]{S. PERCIBALLI}
\affiliation{University of Torino, Via P. Giuria 1, I-10125 Torino, Italy}
\affiliation{INFN-Sezione di Torino, Via P. Giuria 1, I-10125 Torino, Italy}

\author{F. PERFETTO}
\affiliation{INFN-Sezione di Napoli, Via Cintia, I-80126, Napoli, Italy}

\author{A. PERINELLI}
\affiliation{University of Trento, Via Sommarive 14, I-38123 Povo (Trento), Italy}
\affiliation{INFN-TIFPA, Via Sommarive 14, I-38123 Povo (Trento), Italy}

\author[0000-0002-7986-3321]{P. PICOZZA}
\affiliation{University of Roma ``Tor Vergata'', Via della Ricerca Scientifica 1, I-00133 Roma, Italy}
\affiliation{INFN-Sezione di Roma ``Tor Vergata'', Via della Ricerca Scientifica 1, I-00133 Roma, Italy}

\author[0000-0001-5207-2944]{M. PIERSANTI}
\affiliation{University of L'Aquila, Via Vetoio, I-67100, L'Aquila, Italy}

\author[0000-0003-0279-5436]{M. POZZATO}
\affiliation{INFN-Sezione di Bologna, Viale Berti Pichat 6/2, Bologna, Italy}

\author[0000-0001-8587-592X]{G. REBUSTINI}
\affiliation{INFN-Sezione di Roma ``Tor Vergata'', Via della Ricerca Scientifica 1, I-00133 Roma, Italy}

\author[0000-0002-9530-6779]{D. RECCHIUTI}
\affiliation{INAF-IAPS, Via Fosso del Cavaliere 100, I-00133, Roma, Italy}

\author[0000-0002-4222-9976]{E. RICCI}
\affiliation{University of Trento, Via Sommarive 14, I-38123 Povo (Trento), Italy}
\affiliation{INFN-TIFPA, Via Sommarive 14, I-38123 Povo (Trento), Italy}

\author[0000-0001-6816-4894]{M. RICCI}
\affiliation{INFN-LNF, V. E. Fermi, 54, I-00044 Frascati (Roma), Italy}

\author[0000-0001-6176-3368]{S.~B. RICCIARINI}
\affiliation{IFAC-CNR, Via Madonna del Piano, 10, I-50019 Sesto Fiorentino (Firenze), Italy}

\author[0000-0001-7884-2310]{A. RUSSI}
\affiliation{INAF-IAPS, Via Fosso del Cavaliere 100, I-00133, Roma, Italy}

\author[0000-0003-1176-2003]{Z. SAHNOUN}
\affiliation{University of Bologna, Viale Berti Pichat 6/2, Bologna, Italy}
\affiliation{INFN-Sezione di Bologna, Viale Berti Pichat 6/2, Bologna, Italy}

\author{U. SAVINO}
\affiliation{University of Torino, Via P. Giuria 1, I-10125 Torino, Italy}
\affiliation{INFN-Sezione di Torino, Via P. Giuria 1, I-10125 Torino, Italy}

\author[0000-0003-3253-2805]{V. SCOTTI}
\affiliation{University of Napoli ``Federico II'', Via Cintia, I-80126, Napoli, Italy}
\affiliation{INFN-Sezione di Napoli, Via Cintia, I-80126, Napoli, Italy}

\author{X. SHEN}
\affiliation{National Space Science Center, Chinese Academy of Sciences, Beijing 100190, People's Republic of China}

\author[0000-0002-6314-6117]{R. SPARVOLI}
\affiliation{University of Roma ``Tor Vergata'', Via della Ricerca Scientifica 1, I-00133 Roma, Italy}
\affiliation{INFN-Sezione di Roma ``Tor Vergata'', Via della Ricerca Scientifica 1, I-00133 Roma, Italy}

\author{S. TOFANI}
\affiliation{INAF-IAPS, Via Fosso del Cavaliere 100, I-00133, Roma, Italy}

\author{N. VERTOLLI}
\affiliation{INAF-IAPS, Via Fosso del Cavaliere 100, I-00133, Roma, Italy}

\author{V. VILONA}
\affiliation{INFN-TIFPA, Via Sommarive 14, I-38123 Povo (Trento), Italy}

\author[0000-0001-8040-7852]{V. VITALE}
\affiliation{INFN-Sezione di Roma ``Tor Vergata'', Via della Ricerca Scientifica 1, I-00133 Roma, Italy}

\author{U. ZANNONI}
\affiliation{INAF-IAPS, Via Fosso del Cavaliere 100, I-00133, Roma, Italy}

\author{Z. ZEREN}
\affiliation{National Institute of Natural Hazards, Ministry of Emergency Management of China, Beijing 100085, People's Republic of China}

\author{S. ZOFFOLI}
\affiliation{Italian Space Agency, Via del Politecnico, I-00133 Roma, Italy}

\author[0000-0001-6132-754X]{P. ZUCCON}
\affiliation{University of Trento, Via Sommarive 14, I-38123 Povo (Trento), Italy}
\affiliation{INFN-TIFPA, Via Sommarive 14, I-38123 Povo (Trento), Italy}

\begin{abstract}
%limit: 250 words
In this paper we report the detection of five strong Gamma-Ray Bursts (GRBs) by the High-Energy Particle Detector (HEPD-01) mounted on board the China Seismo-Electromagnetic Satellite (CSES-01), operational since 2018 on a Sun-synchronous polar orbit at a $\sim$ 507 km altitude and 97$^\circ$ inclination. HEPD-01 was designed to detect high-energy electrons in the energy range 3 - 100 MeV, protons in the range 30 - 300 MeV, and light nuclei in the range 30 - 300 MeV/n. Nonetheless, Monte Carlo simulations have shown HEPD-01 is sensitive to gamma-ray photons in the energy range 300 keV - 50 MeV, even if with a moderate effective area above $\sim$ 5 MeV. A dedicated time correlation analysis between GRBs reported in literature and signals from a set of HEPD-01 trigger configuration masks has confirmed the anticipated detector sensitivity to high-energy photons. A comparison between the simultaneous time profiles of HEPD-01 electron fluxes and photons from GRB190114C, GRB190305A, GRB190928A, GRB200826B and GRB211211A has shown a remarkable similarity, in spite of the different energy ranges.
The high-energy response, with peak sensitivity at about 2 MeV,  and moderate effective area of the detector in the actual flight configuration explain why these five GRBs, characterised by a fluence above $\sim$ 3 $\times$ 10$^{-5}$ erg cm$^{-2}$ in the energy interval 300 keV - 50 MeV, have been detected. 
\end{abstract}


\section{Introduction}
\label{sec:intro}

Cosmic impulsive gamma-ray signals, namely Gamma-Ray Bursts (GRBs), are extremely intense electromagnetic events detected at the top of the Earth's atmosphere by high-energy photon detectors. GRBs were serendipity discovered in 1967 by the Vela satellites designed to monitor nuclear explosions during the Cold War \citep{Klebesadel_1973}. Just after their discovery, it was clear that GRBs were among the most energetic phenomena present in the universe, and a number of dedicated instruments were built to systematically study them and understand their complex nature. Important progresses were obtained with the launch of the NASA Gamma Ray Observatory (GRO), carrying on board the Burst And Transient Source Experiment (BATSE) all sky monitor, specifically designed to detect GRBs. For the first time, BATSE provided a wide energy range coverage and high timing resolution \citep{Fishman_2013}, discovering important characteristics of GRBs, such as a bimodal time behavior that divides GRBs into short and long ones, according to their T\textsubscript{90} time duration (either smaller or larger than 2 seconds), see \citet{Mazets_1981, Dezalay_1991, Kouveliotou_1993} and references therein.
T\textsubscript{90} measures, in seconds, the duration of the time interval during which 90\% of the total observed counts have been detected. T\textsubscript{90} starts after 5\% of the total GRB counts have been detected, and ends when the detected GRB counts are 95\%.
It was also clear that GRB flux was normally peaking at keV energies, usually accompanied by low- and high-energy components \citep{Band_1993}. In addition, their isotropic spatial distribution was confirmed, even though GRO was not able to discern whether their nature was galactic or extragalactic. The cosmic origin of GRBs was finally unveiled with the first unambiguous detection of their optical counterpart by the Hubble Space Telescope \citep{Van_Paradjis_1997}, and the discovery of the X-Ray afterglow by the BeppoSAX satellite \citep{Costa_1997}. Following these discoveries, ESA promoted the INTEGRAL mission \citep{winkler_2003}, able to detect GRBs in the energy range 3 keV - 10 MeV and with arc-min localization, accompanied by gamma-ray all-sky monitoring with 85\% efficiency and realtime web GCN (GRB Coordinates Network) alert features \citep{Mereghetti_2003}.\\
Subsequently, NASA approved the Swift satellite program, now renamed Neil Gehrels Swift Observatory \citep{Gehrels_2004} featuring hard X-Ray arc-min angular resolution localization, prompt detection by the BAT soft gamma-ray imager, and unprecedented fast repointing with follow-up imaging capability by the X-Ray Telescope (XRT). Swift ultraviolet coverage by the UVOT telescope helped clarify that GRBs are strong extragalactic explosions, so far detected up to redshift 8.2 \citep{Campana_2022}.
Short and long GRBs are now known to be generated by very different collapsing processes: the long ones are linked to a particular type of core-collapse (fast-rotating) supernova subset \citep{Galama_1998, Cano_2017}. The short ones have been recently confirmed to be generated by binary neutron star mergers, thanks to the simultaneous detection of GW170817 by ground interferometers \citep{Abbott_2017_PhysRevLett} and GRB170817A by Fermi and INTEGRAL Observatories \citep{Goldstein_2017, Abbott_2017, Savchenko_2017, Ubertini_2019}.
Finally, the hypothesis that at least some GRBs could be generated by giant \textit{magnetar} flares \citep{Mazets_1979, Mazets_2008, Svinkin_2021, Burns_2021} is still debated as shown by \citet{Schösser_2023}.\\
A handful of them, the strongest ones, have also shown a gamma-ray afterglow, as also recently confirmed by observations of the so-called BOAT (Brightest Of All Time), also known as GRB221009A (see \citet{Burns_2023, Rodi_2023} and references therein). Also, this GRB has been observed at energy greater than 10 TeV by the LHAASO detector with an operative energy range up to 18 TeV \citep{Huang_2022}.\\
GRBs are usually detected using electromagnetic detectors. However, upon the occurrence of GRB221009A, we were able to identify a mechanism which, through the conversions of X and gamma rays, allowed solid state  particle sensors, such as the HEPP-L detector also flying on board the CSES-01 satellite \citep{Shen_2018}, to accurately detect the most intense part of the BOAT GRB \citep{Battiston_2023}. 
The present study is meant to extend this latter analysis to HEPD-01 data, with the aim to provide additional information on the reported GRBs, and in particular to provide the light curve and fluence in the energy range 300 keV - 50 MeV.

\section{CSES-01 mission and the HEPD-01 detector} 

\begin{figure}[h!]
\plotone{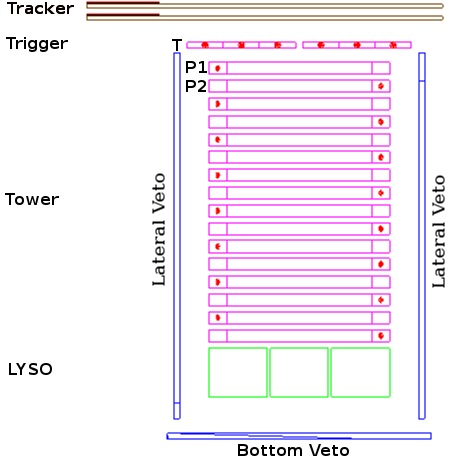}
\caption{Schematic of the HEPD-01 detector. All mechanical structures (as well as the lateral VETO planes located in the front and in the back) have been removed from the figure for visualization purposes.}
\label{fig:HEPD-01}
\end{figure}

CSES-01 is the first item of an extended constellation of Low-Earth Orbit (LEO) satellites, designed for monitoring perturbations of electromagnetic fields and waves, plasma and charged particle fluxes induced by both natural (earthquakes, solar events, cosmic rays, etc.) and anthropogenic sources in the near-Earth space. For these purposes, since February 2, 2018, CSES-01 has been flying on a Sun-synchronous polar orbit at a $\sim$ 507 km altitude, a 97$^\circ$ inclination, and a $\sim$ 5 day revisiting periodicity.\\
CSES-01 relies on nine instruments operating between $\pm$70$^\circ$ latitude, since they are usually switched off beyond these values due to adjustments in attitude and additional scheduled maneuvers. One of these payloads is the High-Energy Particle Detector (HEPD-01), which was designed and built by the Limadou Collaboration, the Italian portion of the CSES mission. A schematic representation of this compact (40.36 $\times$ 53.00 $\times$ 38.15 cm$^3$) and light ($\sim$ 45 kg) apparatus is depicted in Figure \ref{fig:HEPD-01}.
From top to bottom, HEPD-01 is equipped with the following set of sub-detectors: a tracker made up of two double-sided silicon microstrip planes (213.2 $\times$ 214.8 $\times$ 0.3 mm$^3$), a trigger system (T) including one EJ-200 plastic scintillator layer segmented into six paddles (20 $\times$ 3 $\times$ 0.5 cm$^3$ each), a range calorimeter comprising a tower of 16 plastic scintillator planes, P1, P2, ..., P16 (15 $\times$ 15 $\times$ 1 cm$^3$) and a matrix of 3 $\times$ 3 lutetium–yttrium oxyorthosilicate (LYSO) inorganic scintillator crystals (5 $\times$ 5 $\times$ 4 cm$^3$). Finally, the detector is completed by an anti-coincidence (VETO) system composed of 5 plastic scintillator planes, out of which 4 are placed at the lateral sides of the apparatus and 1 at the bottom. 
The combination of these sub-detectors optimizes the detection of electrons and protons in the 3 - 100 MeV and $\sim$ 30 - 300 MeV energy ranges, respectively, as well as the measurement of light nuclei, with a $\pm$60$^\circ$ field of view and a geometrical acceptance of more than 400 cm$^2$ sr.\\
Since the end of the commissioning phase in August 2018, HEPD-01 has been returning valuable information on galactic protons \citep{Bartocci_2020} and their solar modulation \citep{Martucci_2023_ApJL}, trapped protons in the South Atlantic Anomaly \citep{Martucci_2022}, and space weather phenomena, such as geomagnetic storms \citep{Palma_2021, Piersanti_2022} and solar energetic particle events \citep{Martucci_2023_SpWea}. A more detailed description of the HEPD-01 detector can be found in \citet{Picozza_2019}, \citet{Ambrosi_2020} and in \citet{Ambrosi_2021}.

ass\section{HEPD-01 Data Acquisition} 
\label{sec:methods}

HEPD-01 spends most of CSES-01 orbit time performing acquisition runs, during which scientific data are acquired only for incoming particles satisfying a single predefined trigger mask configuration. A dedicated command can be transmitted in order to set one of the eight predefined trigger mask configurations \citep{Sotgiu_2021}, which are given by different logic combinations of counters from the various sub-detectors. Hence, the different trigger masks define the aperture and the energy acceptance of the instrument. For example, the upmost trigger condition, labeled as T, corresponds to an above-threshold signal only in the trigger plane, and it is associated with the lowest energy threshold. By requiring a deeper penetration of the particle inside the detector (i.e., using the trigger plane counters and a set of tower planes in ``AND'' configuration, such as T \& P1, T \& P1 \& P2, and so on), the geometric factor of HEPD-01 decreases, and consequently, the energy threshold for triggering increases. In case of electron detection, the energy thresholds resulting from T, T \& P1 and T \& P1 \& P2 trigger conditions are $>$ 3 MeV, $>$ 4.5 MeV, and $>$ 8 MeV, respectively. Moreover, each of these predefined trigger masks can be used with different VETO settings: no veto, lateral veto alone, bottom veto alone, whole veto (lateral$+$bottom). At the end of the commissioning phase, following six months of calibration and testing, HEPD-01 was configured with a trigger condition, labeled as T \& P1 \& P2, which corresponds to event acquisition and processing only when the deposited signals in the trigger plane and the first two calorimeter planes (P1, P2) are above predefined thresholds, consequently constraining the energy of the incoming particles processed during data acquisition. 
However, for each of the eight predefined masks, even when not selected for the online acquisition, a rate meter independently provides the corresponding trigger counting rate with 1-second resolution. For each event passing the online trigger condition, the rate meters of each trigger mask are transmitted along with other scientific data, thus allowing the off-line analysis for different particle selections and providing an estimation of the energy and angular dependencies.\\
As for previous space weather studies \citep{Palma_2021, Piersanti_2022}, these trigger-configuration rate meters have proven extremely helpful for GRB detection, since the online high-energy threshold configuration, T \& P1 \& P2, is useless due to the very low-energy particles involved in such phenomena, whereas the rate meter of a less deep configuration (such as T) exhibits a clear and significant response to transients of various physical nature.

\section{HEPD-01 as a gamma-ray detector} 
\label{sec:methods}

\begin{figure}[h!]
\includegraphics[width=\columnwidth]{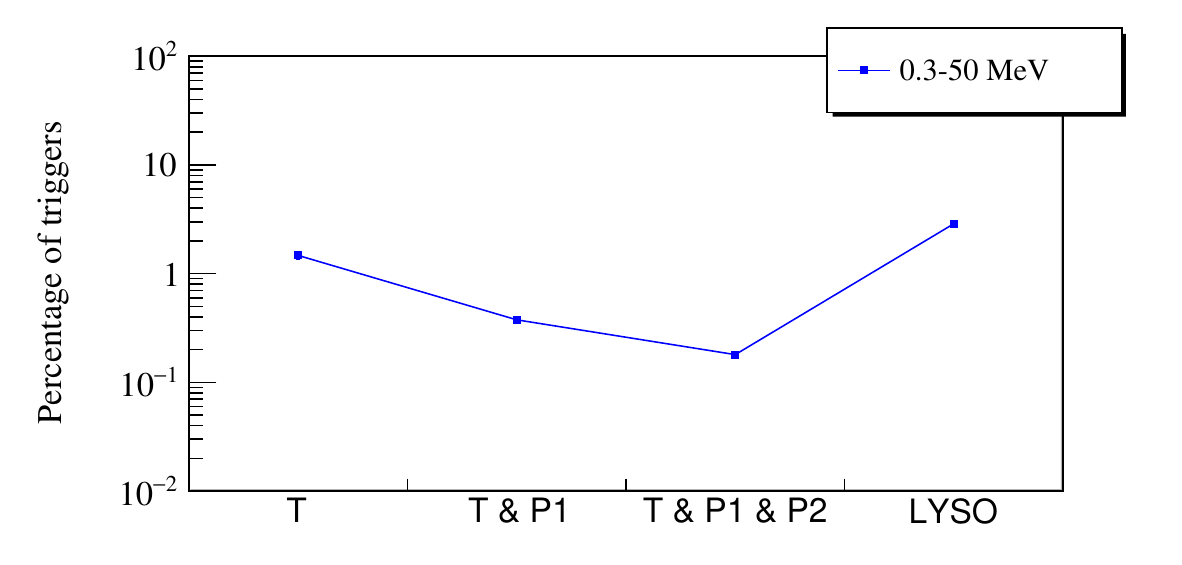}
\caption{Percentage of triggers, estimated from Monte Carlo simulation, as a function of trigger configuration in the photon energy interval 300 keV - 50 MeV.}
\label{fig:Geant4_sim}
\end{figure}

\begin{figure}[h!]
\includegraphics[width=\columnwidth]{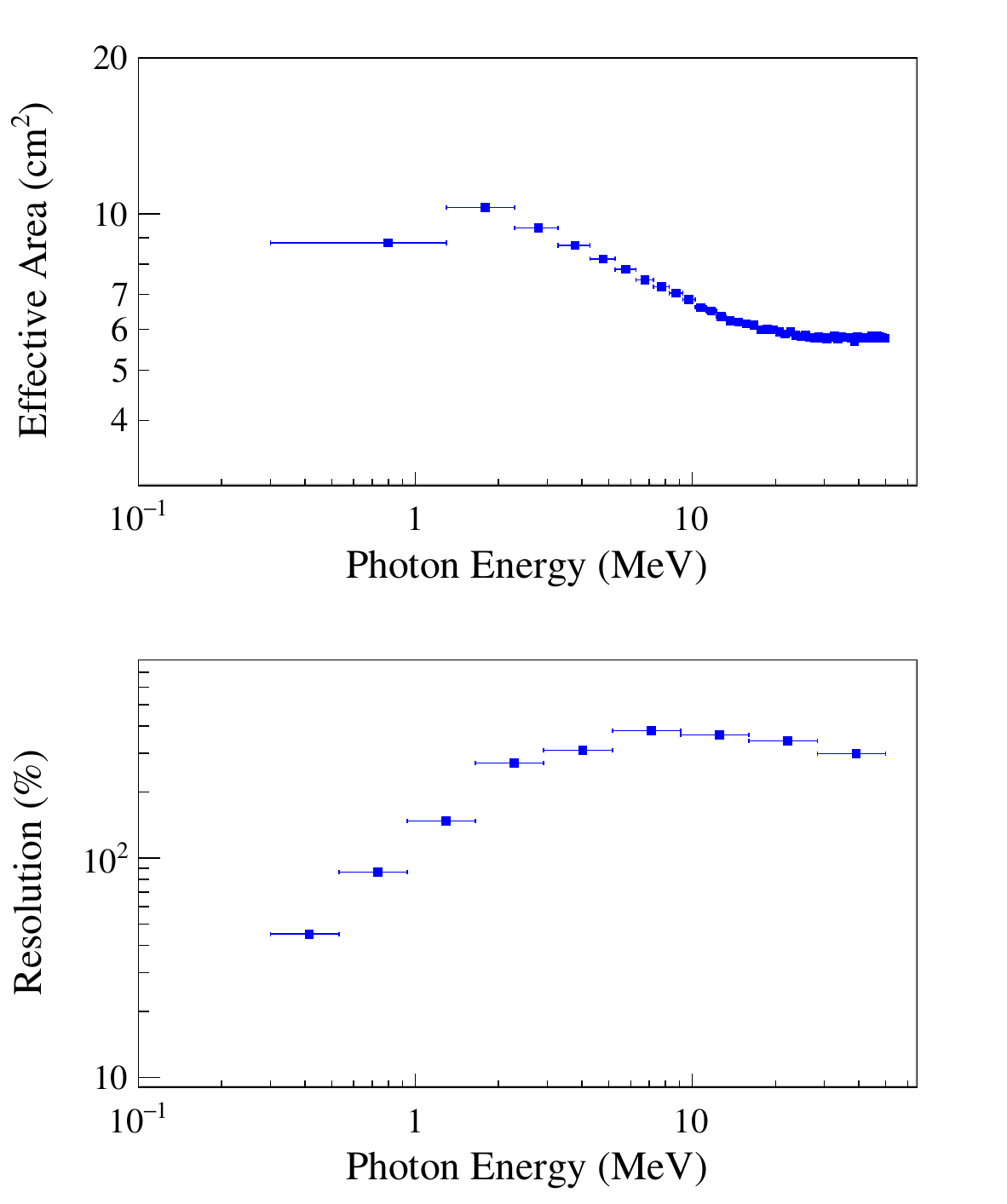}
\caption{On-axis effective area (top panel) and energy resolution (bottom panel) of the HEPD-01 detector in the trigger configuration used for the GRB data analysis.}
\label{fig:Resolution_and_effective_area}
\end{figure}

In order to better assess the HEPD-01 capability to detect high-energy gamma-ray photons, a Monte Carlo simulation was performed using the official software developed by the CSES/Limadou Collaboration and based on a GEANT4 toolkit \citep{Agostinelli_2003}. A sample of 100 million photons was isotropically generated according to a flat distribution in cos$^2\theta$ (0$^\circ < \theta < 90^\circ$, 0$^\circ < \Phi < 180^\circ$), and from a 40 $\times$ 40 cm$^2$ surface immediately above HEPD-01 particle entrance window. Finally, gamma rays were simulated according to a logarithmic energy distribution ranging from 300 keV to 50 MeV, in order to study the detector response over a wide energy interval. To estimate the percentage of triggers as a function of the trigger configuration, we counted the occurrences of charged particle formation (produced by the photons themselves) by requiring an energy deposition above threshold for different combinations of sub-detectors. Then, assuming a single interaction for each photon, we calculated the percentage as a ratio between this number and the initial number of gamma particles. Moreover, since LYSO crystals are usually employed as a valid instrument for low-energy electron and gamma detection, in this simulation we included not only the most superficial layers (T, P1 and P2) and their combinations, but also a hypothetical configuration requiring only an above-threshold signal from the LYSO matrix. Indeed, a dedicated trigger mask including only the LYSO sub-detector was not foreseen during the design phase of the trigger board and, thus, it is not currently implemented in HEPD-01. 
Figure~\ref{fig:Geant4_sim} shows the percentage of triggers decreasing by a factor $\sim$10 when passing from T to T \& P1 \& P2 trigger mask.
Indeed, as observed from in-flight rate meters, the trigger configurations, requiring the coincidence among above-threshold signals from the trigger plane and an increasing number of tower planes, are less and less efficient in detecting the low-energy secondary electrons, which are produced by the photon interactions. Conversely, even after passing through all the 16 tower planes, a significant fraction of photons interacts in the dense high-Z LYSO cubes, as clearly observed in Figure~\ref{fig:Geant4_sim}. This confirms the expected high efficiency of the LYSO matrix in detecting electrons and positrons from gamma-ray conversion despite the matrix is located at the bottom of the payload. For this reason, the second High-Energy Particle Detector (HEPD-02) will have a dedicated trigger system for gamma-ray detection through two planes of LYSO orthogonal bars. The Monte Carlo simulation results confirm that the mechanism in place, when high-energy photons illuminate the upmost section of the detector, is their interaction with the active (tracker silicon planes) and passive materials (aluminum structures etc). This includes photo-ionization, Compton effect and pair production, all generating secondary electrons with energies correlated to the primary gamma-rays.
Figure~\ref{fig:Resolution_and_effective_area} shows the instrument on-axis effective area obtained with the trigger configuration used for the GRB data analysis.
The operative energy range is from 300 keV to 50 MeV, with peak efficiency at $\sim$ 2 MeV, and the actual on-axis effective area spans from $\sim$ 10 cm$^{2}$ at lower energy to $\sim$ 6 cm$^{2}$ at high energy. The energy resolution $\Delta$E/E, in the same trigger configuration, is intrinsically moderate due to the fact that the detector was designed to detect charged particles, vetoing photons.\\
The production of low-energy electrons triggered by the photon interaction in the upper sections of the HEPD-01 is clearly visible during the GRB light curve detection. The effect is the enhancement of the count rate recorded by the rate meter corresponding to trigger configuration T, which requires an above-threshold signal only in the trigger plane and allows the detection of the lowest energetic electrons ($>$ 3 MeV). Adding more tower planes  in ``AND'' configuration with the counters of the trigger plane T suppresses the particle rate. This is why the rate meters of deeper trigger masks (such as T \& P1, T \& P1 \& P2, and so on) do not show any increase at GRB time.

\section{Gamma-Ray Burst observations} 
\label{sec:GRB_observations}

This paper reports the clear detection by HEPD-01 of five strong GRBs, via an electron-flux time profile which closely matches the time evolution of GRB photons mostly detected by dedicated gamma-ray detectors, in the energy range 1 - 1000 keV. HEPD-01 detections were limited to events with a fluence $> 10^{-5}$ erg cm$^{-2}$ in the actual flight configuration optimised to reject photon triggers favouring charged particle ones. In Table~\ref{tab:GRB_parameters} we report the fluence derived for each GRB in the HEPD-01 operative range 300 keV - 50 MeV. Errors take into account a 20\% contribution due to systematic uncertainties. The simultaneous light curves from GRB190114C, GRB190305A, GRB190928A, GRB200826B and GRB211211A, reported in the literature, are very similar to the HEPD-01 ones, demonstrating a limited evolution of the spectral shape at higher energies, as will be discussed later.

\begin{deluxetable*}{cccccccc}[ht!]
\tablenum{1}
\tablecaption{Main parameters for each of the five selected GRBs.\label{tab:GRB_parameters}}
\tablewidth{0pt}
\tablehead{
\colhead{GRB identifier} & \colhead{Right ascension} & \colhead{Declination} & \colhead{Trigger time} & 
\colhead{HEPD-01} & \colhead{Konus-Wind} & \colhead{HEPD-01} 
%& \colhead{Peak} 
\\
\colhead{} & \colhead{(deg.)} & \colhead{(deg.)} & \colhead{(UTC)} & \colhead{duration (s)} & \colhead{fluence (erg cm$^{-2}$)} & \colhead{fluence (erg cm$^{-2}$)} 
%& \colhead{(keV)} 
\\
\colhead{} & \colhead{} &
\colhead{} & \colhead{} & 
\colhead{} & \colhead{(0.3 - 50 MeV)} & \colhead{(0.3 - 50 MeV)} & \colhead{} 
}
%\decimalcolnumbers
\startdata
GRB190114C & 56.2 & $-31.8$ & 20:57:03 & 6 & (1.6 $\pm$ 0.1) $\times$ 10$^{-4}$ & (2.2 $\pm$ 0.3) $\times$ 10$^{-4}$ 
%& 815 
\\
GRB190305A & 11.3 & $-50.1$ & 13:05:19 & 3 & (1.2 $\pm$ 0.3) $\times$ 10$^{-4}$ & (1.0 $\pm$ 0.3) $\times$ 10$^{-4}$ 
%& 1313 
\\
GRB190928A & 36.6 & +29.5 & 13:13.48 & 11 & (3.3 $\pm$ 0.2) $\times$ 10$^{-4}$ & (2.6 $\pm$ 0.5) $\times$ 10$^{-3}$ 
%& 591 
\\
GRB200826B & 296.3 & +71.8 & 22:09:42 & 6 & (1.1 $\pm$ 0.1) $\times$ 10$^{-4}$ & (3.0 $\pm$ 0.5) $\times$ 10$^{-4}$ 
%& 337 
\\
GRB211211A & 211.3 & +27.1 & 13:59:09 & 9 & (3.4 $\pm$ 0.1) $\times$ 10$^{-4}$ & (7.9 $\pm$ 0.9) $\times$ 10$^{-4}$ 
%& 750 
\\
\enddata
\end{deluxetable*}

\subsection{Dataset}
\label{subsec:data_set}

The entire HEPD-01 dataset since end of commissioning (August 2018) was analyzed in search for signals time-correlated with GRBs detected by gamma-ray instruments.
We only considered GCNs reporting GRBs during periods when HEPD-01 was in operation and data were not affected by strong solar activity, South Atlantic Anomaly passage, etc.\\
Relevant information on the five selected GRBs is given in Table~\ref{tab:GRB_parameters}. In columns 2 and 3, right ascension and declination, respectively, are the best positions from GCNs and literature; column 4 reports trigger times from GCNs and literature;
column 5 reports the T\textsubscript{0}-T\textsubscript{90} duration from the HEPD-01 light curve. The recorded durations are lower than the ones reported in the GCNs for two reasons: i) the bursts at high energy are usually shorter than at low energy; ii) due to the moderate sensitivity of HEPD-01, the instrument triggers well above the continue level of the burst.\\
Details about each burst are reported in the following subsections.

\subsection{GRB190114C}

This GRB was the first detected at sub-TeV energy by MAGIC during prompt and afterglow phases \citep{MAGIC_Collaboration_2019}, after first report at T\textsubscript{0} = 20:57:03 UTC by Swift-BAT \citep{Gropp_2019} in the 15 - 350 keV band (T\textsubscript{90} = 362 s), and by Fermi/GBM \citep{Hamburg_2019} in the 50 - 300 keV range (T\textsubscript{90} = 116 s). The 10 - 1000 keV fluence reported by Fermi was 4.433 $\times$ 10$^{-4}$ erg cm$^{-2}$ \citep{Ajello_2020}. The redshift of the GRB is z = 0.42 \citep{Selsing_2019}, which was used to estimate E\textsubscript{iso} = 2.5 $\times$ 10$^{53}$ erg, where E\textsubscript{iso} is the energy emitted in gamma-rays, assuming that the emission is isotropic. Konus-Wind reported a 20 keV - 10 MeV fluence of (4.83 $\pm$ 0.10) $\times$ 10$^{-4}$ erg cm$^{-2}$ \citep{Frederiks_2019b}.
Further information on the burst characteristics at optical wavelength is given in \citet{Melandri_2022} via the monitoring between 1.3 and 370 days after trigger, which revealed a supernova component (SN 2019jrj), underlying the GRB emission. 
The afterglow emission from this GRB was detected at various wavebands from 0.65 GHz to 23 GeV.

\subsection{GRB190305A} 
 
This bright and long-duration GRB was first reported by AGILE \citep{Ursi_2019a} at T\textsubscript{0} = 13:05:19.34 UTC, followed by Swift \citep{Evans_2019}, MAXI \citep{Nahakira_2019}, Insight-HXMT \citep{Xiao_2019}, IPN triangulation \citep{Svinkin_GCN_2021}, and then Konus-Wind \citep{Kozlova_2019}. The fluence, as reported by Konus Wind, was 1.47 $\times$ 10$^{-4}$ erg cm$^{-2}$.

\subsection{GRB190928A}

This bright and long GRB was first reported by AGILE \citep{Ursi_2019b} followed by the IPN triangulation \citep{Hurley_2019} and CALET \citep{Pal'Shin_2019}. The light curve showed multi-peaked pulses with periods of low emission. It was also detected by Konus-Wind \citep{Frederiks_2019} for a total of about 195 s up to $\sim$ 15 MeV at fluence of 4.9 $\times$ 10$^{-4}$ erg cm$^{-2}$, including a 256-ms peak detected at T\textsubscript{0} + 103 s with a fluence of 9 $\times$ 10$^{-5}$ erg cm$^{-2}$ and peak energy of 591 keV. The burst was also reported by Insight-HXMT \citep{Luo_2019} and AstroSat \citep{Gaikwad_2019}.  

\subsection{GRB200826B}

First reported by the Fermi/GBM team \citep{Malacaria_2020} as an exceptionally bright and long GRB with T\textsubscript{90} = 7.4 s in the 50 - 300 keV band, peak energy of 410 keV and fluence of 1.4 $\times$ 10$^{-4}$ erg cm$^{-2}$. The reported 20 keV - 10 MeV fluence from Konus-Wind was (2.0 $\pm$ 0.1) $\times$ 10$^{-4}$ erg cm$^{-2}$ \citep{Ridnaia_2020}.
An IPN triangulation by Konus-Wind, INTEGRAL and Mars-Odyssey was reported in \citet{Hurley_2020}. 
Konus-Wind \citep{Ridnaia_2020} showed a light curve with bright multi-peak pulses of total duration of 23 s and weak emission detected up to T\textsubscript{0} + 250 s and 10 MeV, with fluence of 2 $\times$ 
10$^{-4}$ erg cm$^{-2}$ and peak energy of 337 keV.

\begin{figure*}[ht!]
\begin{center}$
\begin{array}{cc}
\includegraphics[width=\columnwidth]{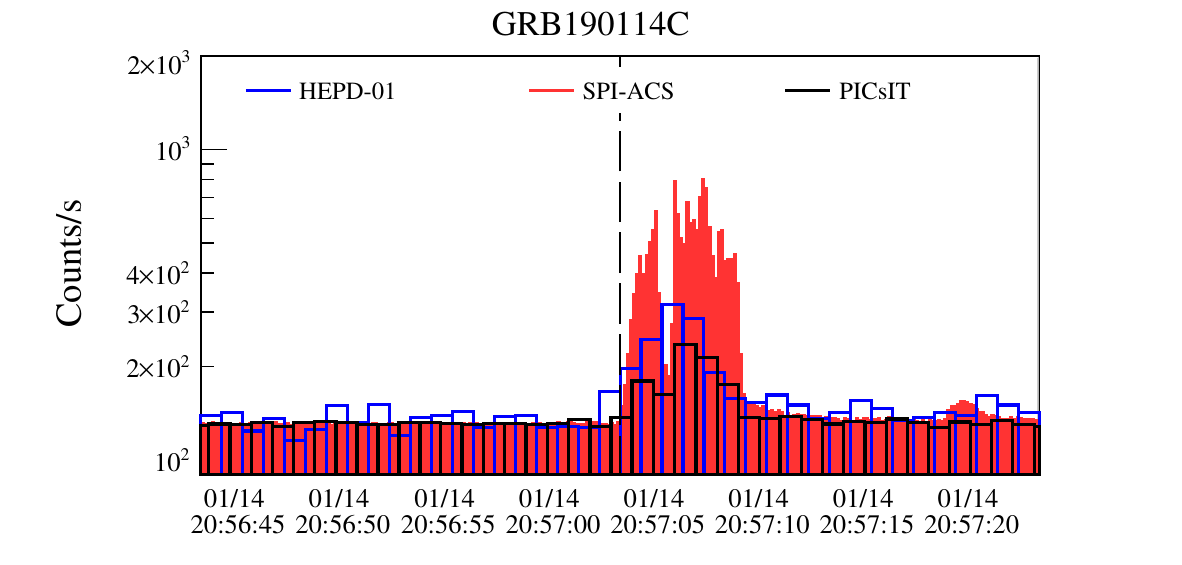}&
\includegraphics[width=\columnwidth]{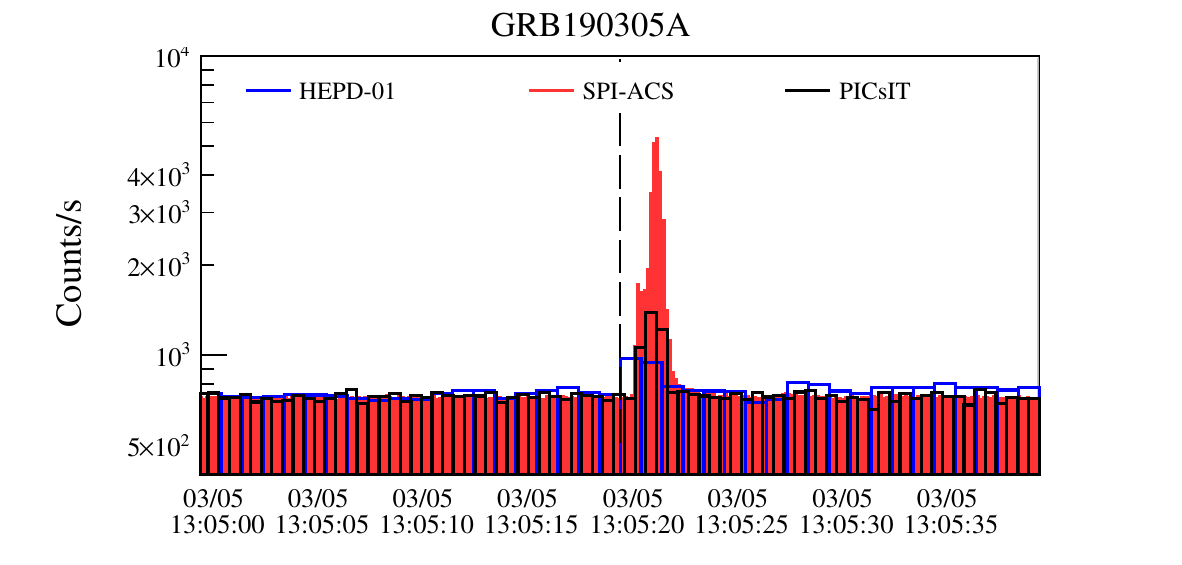}
\end{array}$
\end{center}

\begin{center}$
\begin{array}{cc}
\includegraphics[width=\columnwidth]{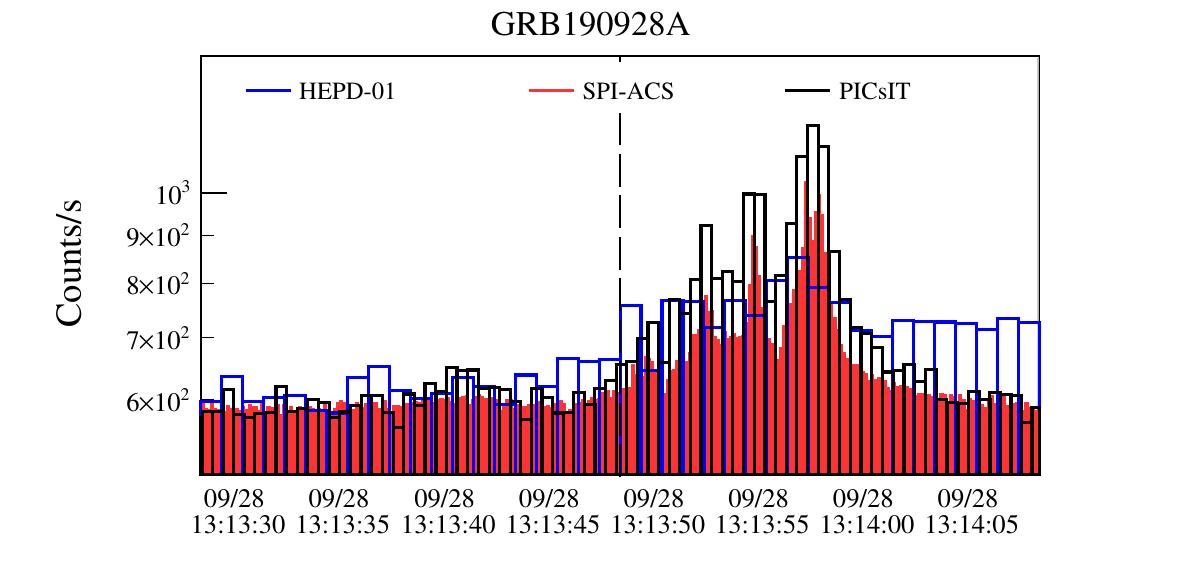}&
\includegraphics[width=\columnwidth]{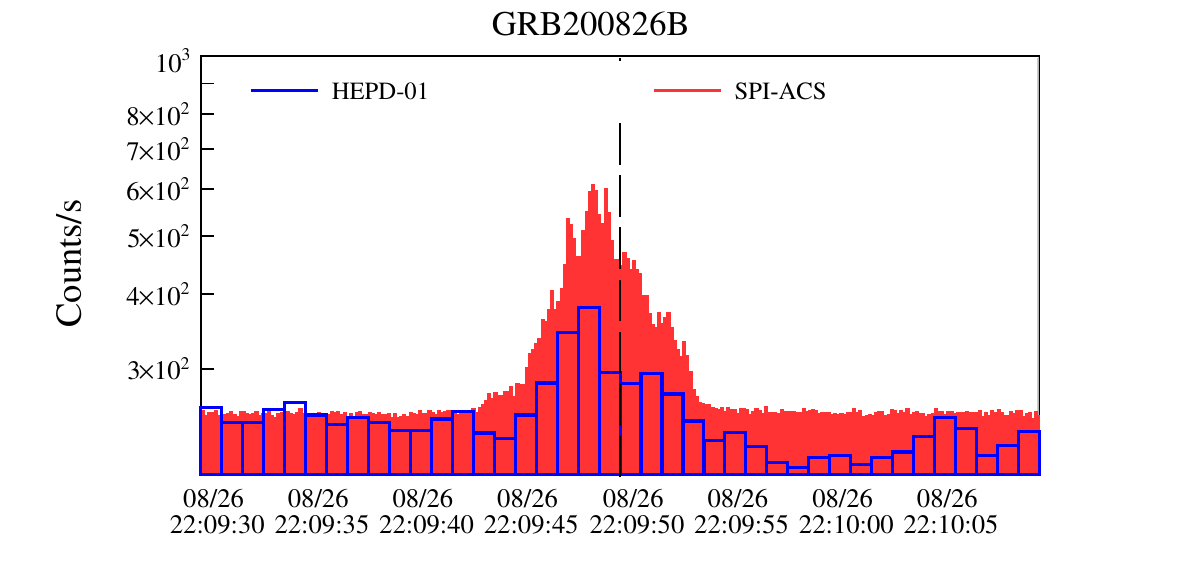}
\end{array}$
\end{center}

\begin{center}$
\begin{array}{c}
\includegraphics[width=\columnwidth]{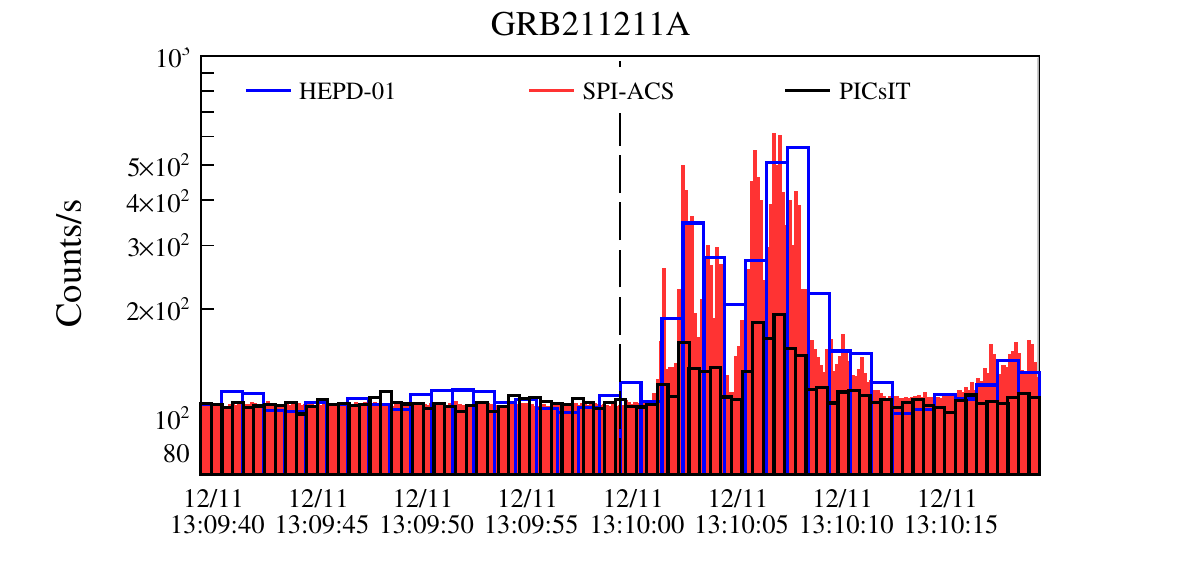}
\end{array}$
\end{center}
\caption{Time-profiles of the low-energy electrons detected by the most superficial trigger configuration of HEPD-01 - in blue - in comparison with the time-profiles of the signals obtained from other instruments like PICsIT and SPI-ACS, in black and red, respectively. For a better visualization, SPI-ACS and PICsIT data points are normalized to HEPD-01 ones. Each panel refers to a specific GRB. The vertical dashed line marks the trigger time T\textsubscript{0} for the start of each GRB.}
\label{fig:GRB_HEPD}
\end{figure*}

\subsection{GRB211211A}

This enigmatic, hybrid GRB was reported by Swift as the second brightest and most complex up to that time \citep{D'Ai_2021} with emission extending up to T\textsubscript{0} + 100 s. Optical afterglow was promptly detected by \citet{Zheng_Filippenko_2021}. A Swift fluence of 1.5 $\times$ 10$^{-4}$ erg cm$^{-2}$ (15 - 150 keV) was then reported in \citet{Stamatikos_2021}. 
Fermi/GBM showed a low-energy curve in the 10 - 300 keV range characterized by 3 separate pulses and peak energy of 646.8 keV. The event fluence is 5.4 $\times$ 10$^{-4}$ erg cm$^{-2}$ (10 - 1000 keV) from T\textsubscript{0} - 1.264 s to T\textsubscript{0} + 54.033 s \citep{Mangan_2021}.
The optical, ultraviolet and X-ray counterpart was immediately searched and finally a nearby host galaxy at distance of 346 Mpc was identified by different authors. 
Further observations confirmed the initial GRB location and reported a peculiar slow decay associated to such long GRB type \citep{Malesani_2021_GCN_31221}. 
In \citet{Malesani_2021_GCN_31223}, Malesani corrected the E\textsubscript{iso} value to 6.9 $\times$ 10$^{51}$ erg. The collected result from observational campaigns made clear the unusual nature of GRB211211A compared to the classical long GRBs at such a distance, and suggested a generation mechanism from a compact binary merger (such as two neutron stars) in order to match location, rapid decay and color (see \citet{Levan_2021}). The features of this event impacted the standard GRB scenario; characteristics, light curves, and new models are reported in \citet{Troja_2022}, \citet{Rastinejad_2022},\citet{Mei_2022} and in \citet{Yang_2022}.

\subsection{GRB time-profiles and detectability}

In Figure~\ref{fig:GRB_HEPD} the time-profiles of the low-energy ($>$ 3 MeV) electrons produced by the GRB impinging photons and detected by the most superficial trigger configuration of HEPD-01, in blue, are compared with the time-profiles of the signals obtained from dedicated gamma instruments.
In this specific case we used data from INTEGRAL/PICsIT \citep{Ubertini_2003, Labanti_2003} and INTEGRAL/SPI-ACS \citep{von_Kienlin_2003}. For a better visualization, SPI-ACS and PICsIT data points are normalized to HEPD-01 ones. The vertical dashed line appearing in each panel marks the T\textsubscript{0} trigger time reported by GCN for the start of the corresponding GRB. Although instruments like SPI-ACS and PICsIT have a higher resolution and are designed for this type of detection, the signals recorded by HEPD-01 show a good agreement in both shape and duration with previously reported observations, also providing a good enough timing resolution if compared to the one provided by PICsIT (7.9 or 15.8 ms depending on the observation period).
\begin{figure*}[ht!]
\begin{center}$
\begin{array}{cc}
\includegraphics[width=\columnwidth]{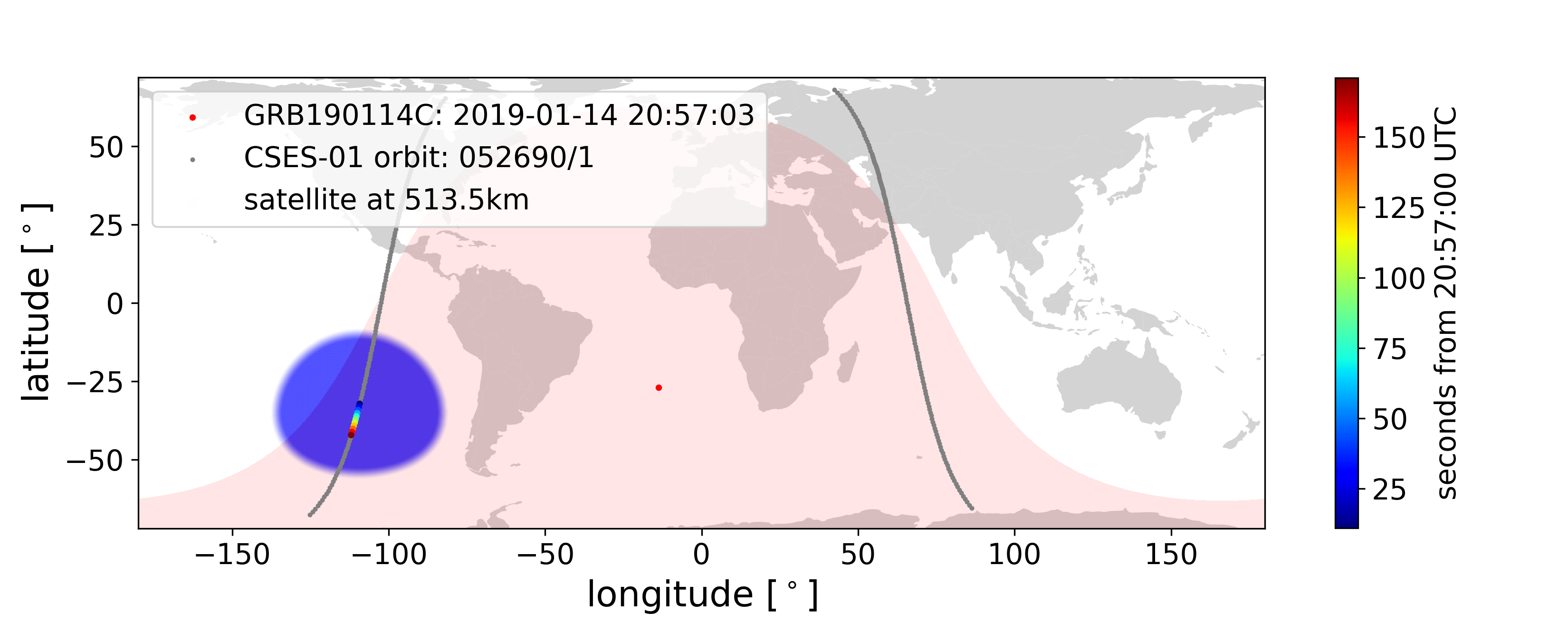}&
\includegraphics[width=\columnwidth]
{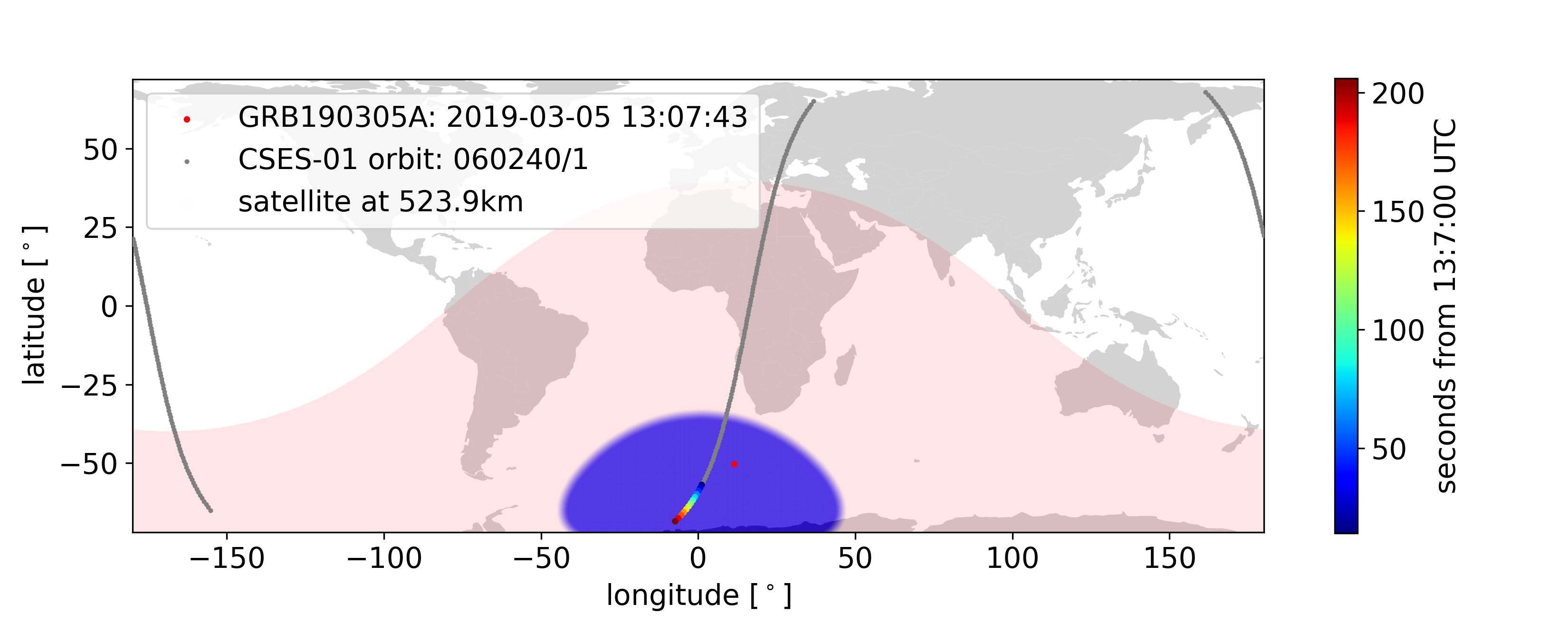}
\end{array}$
\end{center}

\begin{center}$
\begin{array}{cc}
\includegraphics[width=\columnwidth]{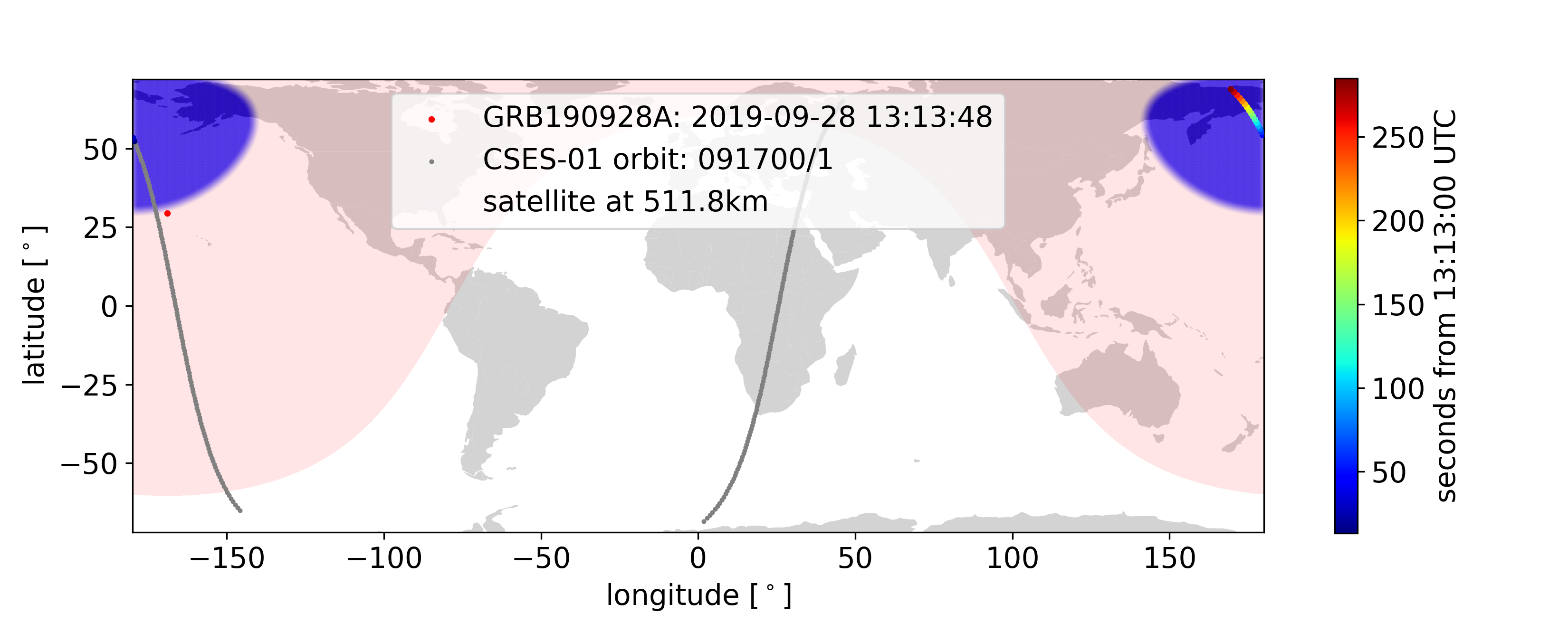}&
\includegraphics[width=\columnwidth]{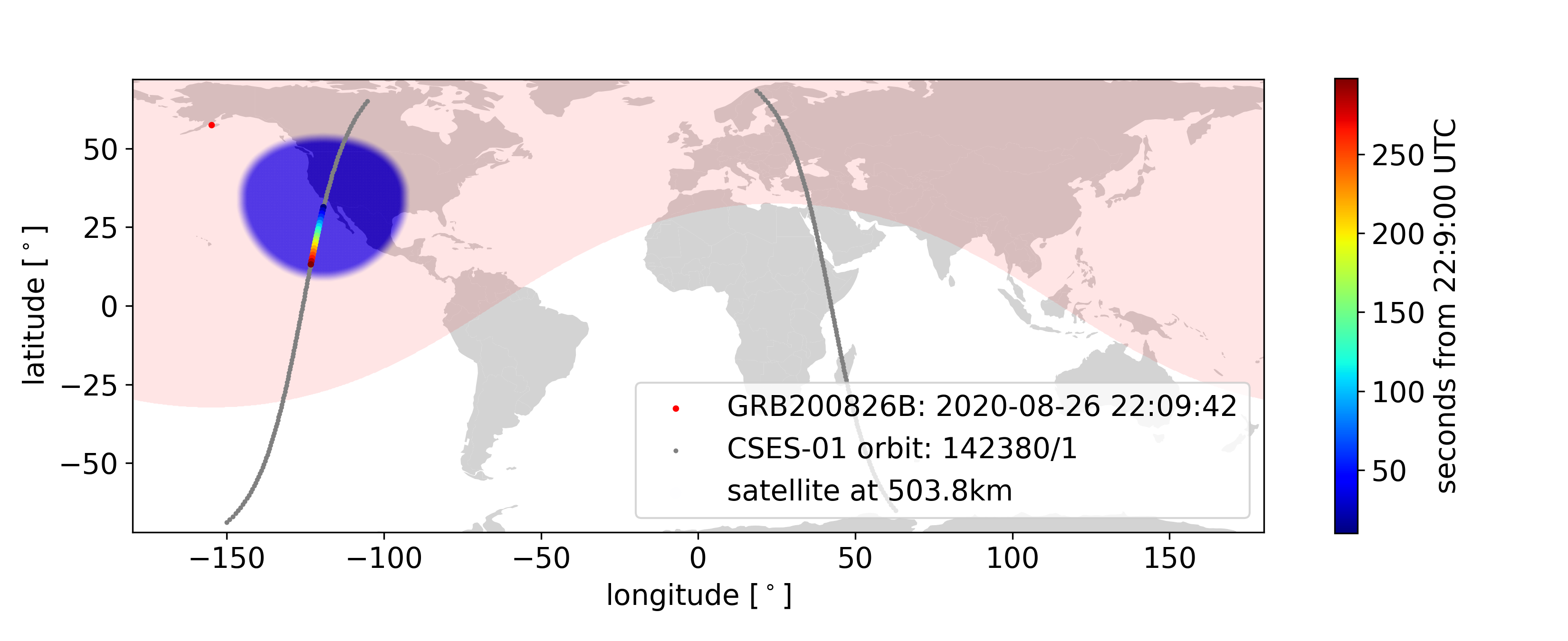}
\end{array}$
\end{center}

\begin{center}$
\begin{array}{c}
\includegraphics[width=\columnwidth]{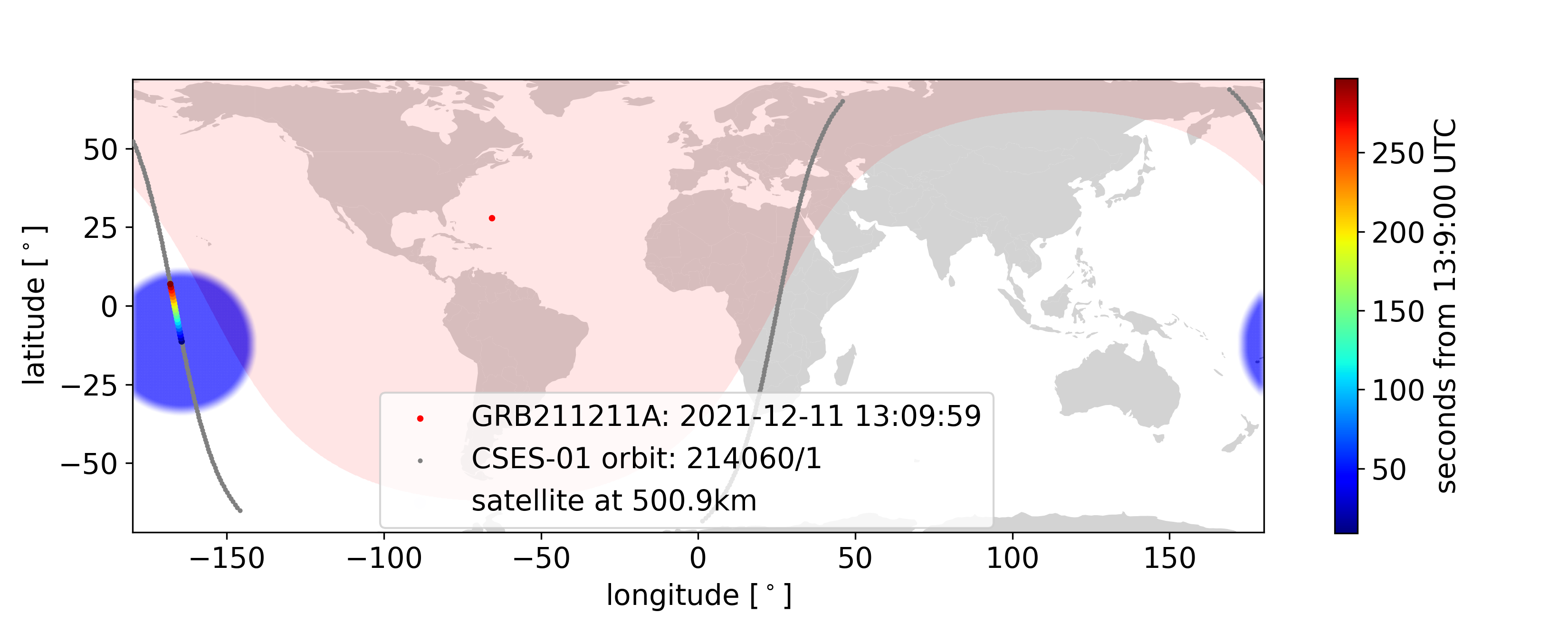}
\end{array}$
\end{center}
\caption{Five Earth maps (in correspondence with the observations shown in Figure~\ref{fig:GRB_HEPD}) which show: the illuminated half-globe of the GRB (red area), the CSES-01 orbits close by in time (grey lines), and the horizon seen at the satellite altitude (given in the legend and shown as a blue area) at the moment of 
first impact of any GRB. The color bar reflects the time in seconds after the first GRB trigger and the corresponding position along the orbit.}
\label{fig:CSES_orbit_maps}
\end{figure*}
Moreover, for each observed GRB, we estimated the significance of the detected signal over background: 50.7$\sigma$ (GRB190114C), 17.6$\sigma$ (GRB190305A), 83.8$\sigma$ (GRB190928A), 24.9$\sigma$ (GRB200826B), and 167.6$\sigma$ (GRB211211A).
We have compared the HEPD-01 (300 keV - 50 MeV) measured fluence with the extrapolated fluence from the Konus-Wind Band model fits (20 keV - 10 MeV). The ratios between the observed fluences and the expected ones are all greater than 1, apart for GRB190305A: 1.4 (GRB190114C), 0.8 (GRB190305A), 7.9 (GRB190928A), 2.7 (GRB200826B), and 2.3 (GRB211211A). The fluences are statistically compatible, within $2\sigma$, for GRB190114C and GRB190305A, while GRB190928A, GRB200826B and GRB211211A are incompatible with the extrapolation of the Konus-Wind spectra values. The higher fluxes could be due to a high-energy spectral component above 10 MeV, detected in GRB190928A, GRB200826B and GRB211211A, as also observed in GRB190114C up to $\sim$ 100 MeV \citep{Ursi_2020}. 
Finally, it is worth noting that the fluence of GRB190928A could be overestimated due to a change in the background rate during the detection, as can be seen in Figure~\ref{fig:GRB_HEPD}. We have taken into account this variation, though there may be a systematic uncertainty larger than in the other cases. \\
Figure~\ref{fig:CSES_orbit_maps} shows the position of the CSES-01 satellite at the moment of the GRBs impact, its illumination area estimated from the right ascension and declination given in Table~\ref{tab:GRB_parameters}, and the satellite's horizon. An overlap between the red and blue area confirms the visibility of the GRBs by the instruments on-board the satellite, which is confirmed for all five GRBs discussed above.

\section{Conclusions} 
\label{sec:conclusions}
The serendipitous detection of five GRBs by the HEPD-01 detector in the energy range 300 keV - 50 MeV has provided additional information on a set of strong events with a typical fluence above $\sim$ 3 $\times$ 10$^{-5}$ erg cm$^{-2}$, which corresponds to the 2$\sigma$ detection limit of HEPD-01 in the trigger configuration running during the reported GRB observation. In particular, the high-energy extension up to 50 MeV, additional coverage and duty cycle thanks to the polar orbit, high time resolution and positional accuracy of the CSES-01 satellite position provide a valuable contribution to the single GRB characterization.\\
The reported results show how a light and compact payload with similar design to HEPD-01's (e.g., HEPD-02) is able to detect extreme astrophysical phenomena like GRBs, and how the presence of LYSO will be crucial for this kind of physics. 
Furthermore, given the fast response of the detector and the instrumental upgrades in HEPD-02, the participation of the mission in a wider, real-time alert program (InterPlanetary Network etc.), is foreseen  during the second mission.\\
The reported GRB observation by HEPD-01 is valuable \textit{per se}, as an independent source of data, given the 5-yr uninterrupted observations, all-sky sensitivity (aside of the Earth obscuration), high-energy extended response, and considering the forthcoming launch of HEPD-02 \citep{DeSantis_2021} on board the CSES-02 satellite. The second-generation HEPD-02 detector has a dedicated trigger system to detect photons interacting with the plastic scintillator range calorimeter, providing a peak sensitive area of $\sim$ 30 cm$^{2}$ at 2 MeV \citep{Lega_2023}, and with the LYSO bulk scintillators featuring an area of $\sim$ 150 cm$^{2}$ between 5 and 50 MeV \citep{Follega_2023}.

\section{acklowledgements}
\begin{acknowledgments}
This work makes use of data from the CSES mission, a project funded by China National Space Administration (CNSA), China Earthquake Administration (CEA) in collaboration with the Italian Space Agency (ASI), National Institute for Nuclear Physics (INFN), Institute for Applied Physics (IFAC-CNR), and Institute for Space Astrophysics and Planetology (INAF-IAPS). This work was supported by the Italian Space Agency in the framework of the “Accordo Attuativo 2020-32.HH.0 Limadou Scienza+” (CUP F19C20000110005), the ASI-INFN Agreement n. 2014-037-R.0, addendum 2014-037-R-1-2017, and the ASI-INFN Agreement n. 2021-43-HH.0. AB, JCR, and PU acknowledge the Italian Space Agency for the financial support under the “INTEGRAL ASI-INAF” agreement n. 2019-35-HH.0. 
\end{acknowledgments}

\bibliography{sample631}{}
\bibliographystyle{aasjournal}

%\bibitem[Fishman et al. (19xx)]{fishman19xx1973} GRO BATSE



\end{document}